\documentclass[aps,twocolumn,pra,showpacs]{revtex4}
\usepackage[T1]{fontenc}
\usepackage[dvips]{graphicx}
\usepackage[latin1]{inputenc}
\usepackage{psfrag}



\begin{document}

\title{Four-photon Entanglement as Stochastic-signal Correlation}

\author{A. F. Kracklauer}
\affiliation{kracklau@fossi.uni-weimar.de}

\begin{abstract}
A fully classical model of a recent experiment exhibiting what is interpreted
as teleportation and four-photon entanglement is described. It is shown that
the reason that a classical model is possible, contrary to the current belief,
results ultimately from a misguided modification by Bohm of the EPR \emph{Gedanken}
experiment. Finally, teleportation is reinterpreted as the passive filtration
of correlated but stochastic events in stead of the active transfer of either
material or information.
\end{abstract}


\maketitle In a recent letter J.-W. Pan et al. described a demonstration of the
experimental observation ``of pure four-photon GHZ entanglement produced by
parametric down-conversion and a projective measurement.''\cite{JWPan} They add
that this experiment ``demonstrates teleportation with very high purity''
and that ``the high visibility not only enables various novel tests of quantum
nonlocality, {[}but{]} it also opens the possibility to experimentally
investigate quantum computation and communications schemes with linear
optics.'' This demonstration was achieved using a novel and ingeniously simple
(in concept, not necessarily in concrete realization) setup employing two
sources of entangled photons which feed the two faces of a polarizing beam
splitter (PBS). The authors further claim that this setup gives results that
finally preclude all doubt that nonlocal effects ensue from quantum
entanglement.

It is the purpose of this letter to contest this last claim. This will be
achieved by proposing a classical model of their experiment, without
`nonlocality,' which fully duplicates its results. 

\begin{figure}[ht]
\begin{center}
\includegraphics[]{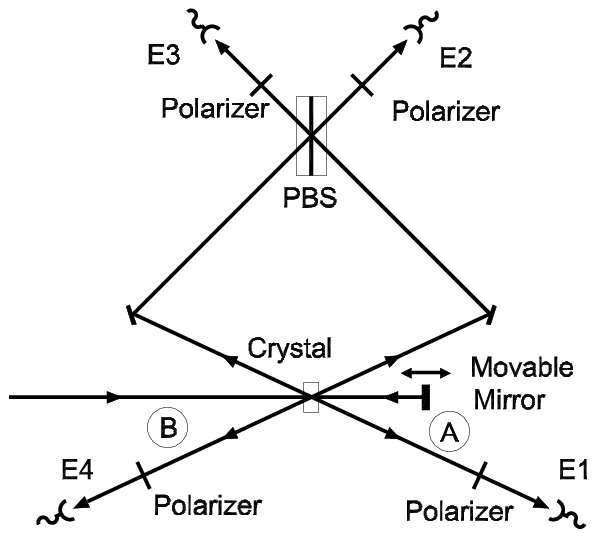}
\end{center}
\caption{Schematic of the experimental setup for the measurement
of four-photon GHZ correlations. A pulse of laser light passes a nonlinear
crystal twice to produce two entangled photon pairs via 
parametric down conversion. Coincidences
between all four detectors are used to study the nature of entanglement. }
\label{setup}
\end{figure}

First, we briefly review the experimental setup. (See Fig.~\ref{setup}) Two
independent entangled photon pairs are created by down-conversion in a
crystal pumped by a pulsed laser. The laser pulse passes through the crystal
creating one pair (A), then is reflected off a movable mirror to repass through
the crystal in the opposite direction creating a second pair (B). One photon
from each pair is fed directly through polarizers to photodetectors (photons 1
and 4). The other photons (2 and 3) are directed to opposite faces of a PBS,
(i.e., a beam splitter which reflects vertically and transmits horizontally
polarized photons) after which the exiting photons are sent through variable
polarizers into photodetectors. The path lengths for photons 2 and 3 are
adjusted so as to compensate for the time delay in the creation of the pairs.
By moving the mirror, the compensation can be negated to permit observing the
disappearance of interference caused by lack of simultaneous ``cross-talk''
between channels 2 and 3.

The reported observations are the following: Of all the 16
possible polarizer settings regimes for which \( \theta _{n}=0\) or \(\pi /2\)
w.r.t. the horizontal axis of the PBS, only
\( \{0,\: \pi /2,\: \pi /2,\: 0\} \) and \( \{\pi /2,\: 0,\: 0,\: \pi /2\}
\) yield a (substantial) four-fold coincidence count: \( C \); the regime \(
\{\pi /4,\: \pi /4,\: \pi /4,\: \pi /4\} \) occurs with an intensity \( C/4.2
\) and the regime \( \{\pi /4,\: \pi /4,\: \pi /4,\: -\pi /4\} \) with zero
intensity. Further, both of the latter regimes yield an intensity of \( C/8 \)
when photons 2 and 3 do not overlap.

The model proposed herein, like all others, is based on certain assumptions,
which, being quite different from those 
couched in the notation and vocabulary of QM, must be delineated explicitly.
They consist of the following:

\begin{enumerate} 
\item Electromagnetic radiation is comprised of continuous
waves as described by Maxwell's equations.  
\item All detectors in the optical
region of the electromagnetic spectrum exploit the photoelectric effect. They
convert continuous radiation to an electron current. Electrons are discrete
objects; their generation in a photodetector effectively digitizes the signal
associated with incoming radiation, thereby evoking
the impression that the radiation was itself somehow digitized into units
(photons). The last inference is unwarranted; we can not know what form
the incoming energy actually had, we are restricted to inferring its nature
from the electron current. It is known empirically that photoelectrons are
ejected from a photodetector randomly but in proportion to the energy density,
\( E^2 \), of the incoming signal.  For the model, we take the conversion
efficiency to be \(100\%\).

\item The nonlinear crystal generating signals by parametric down-conversion is
taken to be a ``black box'' which, by virtue of its structure emits randomly
pairs of signal pulses confined to the pure vertical and horizontal
polarization modes. The pairs are anticorrelated with respect to polarization
and each member of a pair is  directed (in the well known manner) into two
intersecting cones; all horizontal emissions are into one cone, all vertical
into the other. The two variants of pairs are taken to be equally likely.
``Entangled'' radiation samples extracted from the two points where the cones
intersect, therefore, constitute a mirror image pair of random sequences of
individual electromagnetic pulses of both pure modes. 

\item The intensity of four-fold coincidence detections among four
photodetectors is calculated using non quantum  coherence theory.\cite{M&W}
Coincident count probabilities, for a system with \( N \), (herein \( N = 4\)
), monitored exit ports are proportional to the single time, multiple location
second order (in intensity) cross correlation, i.e.:

\begin{equation}
\label{e2}
P(\theta_{1},\, \theta_{2},..\theta_{N})=\frac{<\prod
^{N}_{n=1}E^{*}(r_{n},\,\theta_n)\prod ^{1}_{n=N}E(r_{n},\,\theta_n)>}{\prod
^{N}_{n=1}<E^{*}(r_{n})E_{n}(r_{n})>}. \end{equation}
\end{enumerate}
For this model the denominator consists of constants of factors of the form
\((\cos^2(a)+\sin^2(a))\) so that it equals \(1\).

These assumptions, all fully compatible with classical physics,  account for
all of the reported observations. Eq.~(\ref{e2}) was implemented numerically as
follows: The centers are assumed to emit double pulses in opposed directions
which are anticorrelated and confined to the vertical and
horizontal polarization modes; i.e. their polarization vectors are:

\begin{eqnarray}
\label{e3}
A_{1} & = & (cos(n\frac{\pi }{2}),\: sin(n\frac{\pi }{2})),\\
A_{2} & = & (sin(n\frac{\pi }{2}),\: -cos(n\frac{\pi }{2})),\\ 
B_{1} & = & (cos(m\frac{\pi }{2}),\: sin(m\frac{\pi }{2})),\\
B_{2} & = & (sin(m\frac{\pi }{2}),\: -cos(m\frac{\pi }{2})),
\end{eqnarray}
where \( n \) and \( m \) take the values \( 0 \) and \( 1 \) with a flat 
random distribution. The polarizing beam splitter (PBS) is modeled using the
transition matrix for a polarizer,

\begin{equation} \label{e4}
P(\theta )=\left[ \begin{array}{cc}
\cos ^{2}(\theta ) & \cos (\theta )\sin (\theta )\\
\sin (\theta )\cos (\theta ) & \sin ^{2}(\theta )
\end{array}\right] ,
\end{equation}
where \( \theta =\pi /2 \) accounts for a reflection, and \( \theta =0 \) a
transmission. Thus, the field impinging on each of the four detectors is:
\begin{eqnarray}
\label{e6}
E_{1} & = & P(\theta _{1})A_{1},\\
E_{2} & = & P(\theta _{2})(P(0)B_{2}-P(\pi /2)A_{2}),\\
E_{3} & = & P(\theta _{3})(P(0)A_{2}-P(\pi /2)B_{2}),\\
E_{4} & = & P(\theta _{4})B_{1}.
\end{eqnarray}
 
Putting Eqs. (7-10) into Eq. (1) and carrying out the indicated average by
summing on \( n \) and \( m \) and dividing by \(4\) as there are four
combinations, one obtains the expression for the fourth order (in fields)
coincidence correlation as a function of the polarizer settings. The result is
cumbersome but easily evaluated numerically for various regimes, i.e., various
selections of  \( \theta _{n} \). Ideal results compatible to those observed
are obtained.  Slight deviations in the observed data appear to be due to
noise. 

\psfrag{ZWP}[][][0.7]{$\pi/4$}
\psfrag{PZW}[][][0.7]{$\pi/2$}

\begin{figure}[ht]
\begin{center}
\includegraphics[width=0.99\columnwidth]{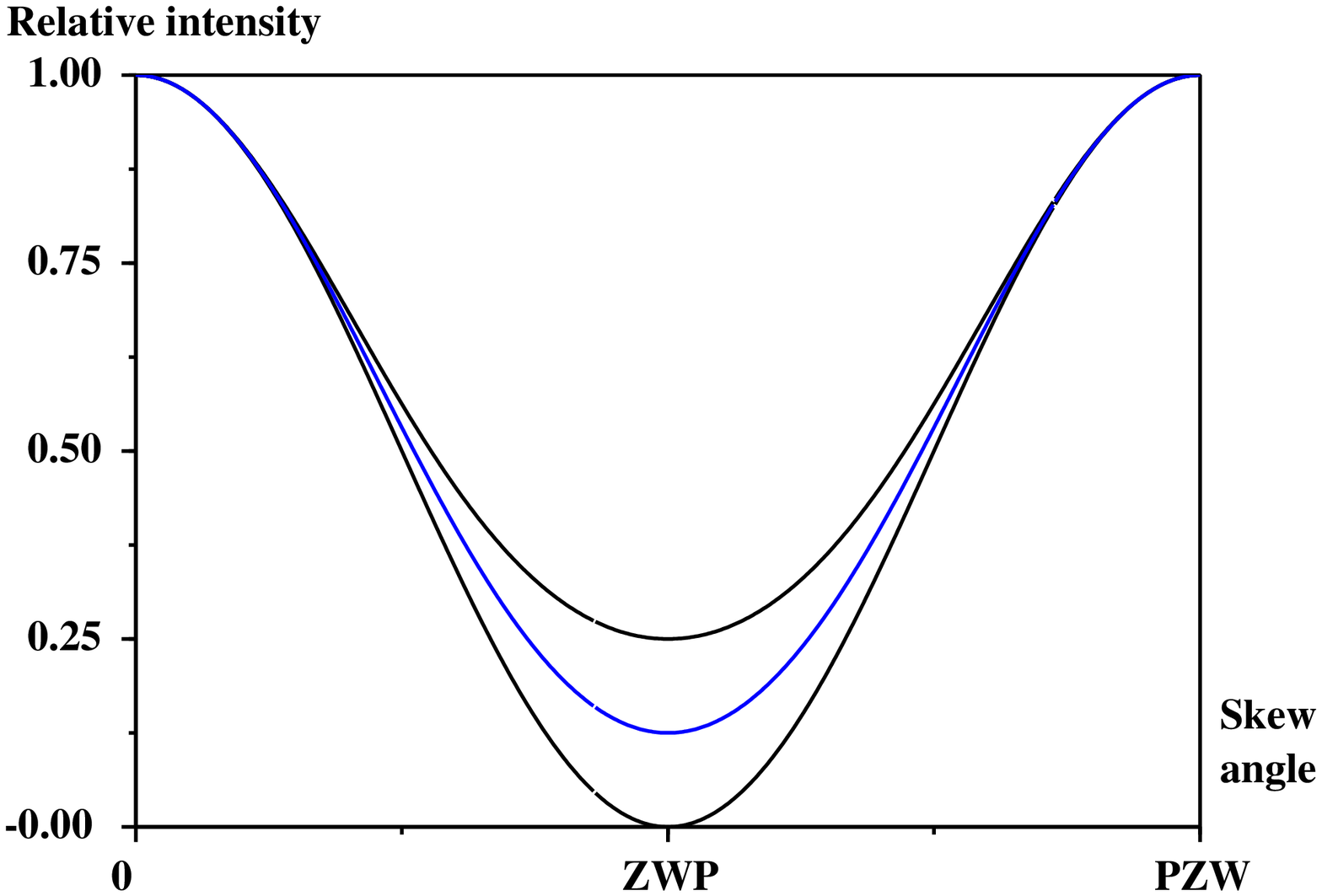}  
\end{center}

\caption{The upper curve shows the effect on the intensity of four-fold 
coincidences of skewing (rotating) all polarizers through a given angle
starting from the state $\{\pi/2,0,0,\pi/2\}$.   The lower  curve shows the
same effect when one of the polarizers is rotated in  the opposite direction.
The middle curve shows the effect of either of  these skewing schemes when the
timing is such that the  crossover signals do not arrive simultaneously with
the reflected signals. Note that the separation of the curves and values at
\(\pi/4 \) coincide with the observations reported in Ref. \cite{JWPan}.  This
diagram differs from Fig. 4 in Ref. \cite{JWPan} in that it shows the split of
these regimes as a function of polarizer skew for fixed delay rather than  as a
function of delay for fixed skew.}

\label{skew}
\end{figure}

To model regimes for which the pulse pairs were generated with a time
difference such that the cross-over signals in beams 2 and 3 could not
interfere, the sum on the radiation contributions from individual centers is
squared \emph{before} averaging. This procedure recognizes the fact that the
electric fields from distinct sources (A and B, in this case) must be added
\emph{after} squaring because they do not interfere but still deposit energy
into a photodetector.  Changing the order of `squaring' and `summing' for this
setup affects only the crossover signals as all others are in orthogonal
polarization modes in any case. All this leads, again, to results exactly
mimicking the effects reported. Additionally, the relative count intensity in
other regimes can be computed easily; for an example, see Fig. \ref{skew}.

Note that on Fig.~2 the curve for the `state' \({<\pi/4,\pi/4,\pi/4,\pi/4>}\)
splits or separates from the `state' \({<\pi/4,\pi/4,\pi/4,-\pi/4>}\) at \(
\theta=\pi/4\). Pan et al. interpret this splitting to indicate: ``that the
state of, say, photon 2 was teleported to photon 4\dots'' and ``\dots fully
demonstrates the nonlocal feature of quantum teleportation.''  The mere fact
that this splitting is faithfully modeled classically, using, e.g., coherent
vice Fock states, casts strong doubt on such inferences.

There is nothing essentially quantum mechanical in this model. It does not make
use of any property peculiar to QM.  In particular it does not in any way
assume properties of ``photons'' and its results do not depend, therefore, on
assumptions regarding detector efficiency (effectively taken to be \(100\%)\). 
It constitutes a fully faithful classical technique for  modeling experiments
carried out in a portion of the spectrum admitting macroscopic devices and
detailed time tracking of electromagnetic fields, thereby evading the
peculiarities of photodetectors. (Note that EPR-B correlations have been so
observed.\cite{Klyshko})  The same type of model for EPR-B  experiments yields
exactly the same expressions as those given by QM which violate the relevant
Bell Inequality.

In view of the conventional wisdom that it is impossible to comprehend
phenomena involving \emph{quantum} entanglement using non quantum physics
\cite{DB'meester}, it is natural to ask how it is possible then that a
classical model can be suitable. We believe that the genesis of this
misunderstanding is to be found with  D. Bohm.\cite{DBohm} The over-arcing
issue ultimately motivating the experiment described in Ref. \cite{JWPan}, and
further analyzed herein is the Einstein, Podolsky and Rosen {[}EPR{]} argument
of 1935 to the effect that QM is incomplete.\cite{EPR} In that paper EPR argued
that the position and momentum of two entangled particles, created
auspiciously, could be specified exactly in spite of Heisenberg Uncertainty, by
exploiting symmetry. With this argument, they hoped to show that Heisenberg
Uncertainty was not something fundamentally new, but just ignorance. This being
the case, they argued, there should exist a deeper theory, involving heretofore
``hidden variables'' that would more precisely describe nature. The point of
their whole consideration involved the ultimate nature of Heisenberg
Uncertainty.

Much later  Bohm modified EPR's original \emph{Gedanken} Experiment. He, for
reasons of simplicity, transfered the EPR argument from phase space, \( (x,\,
p) \) to another arena, one involving spin.\cite{DBohm} This too turned out to
be experimentally unrealizable, but the algebraically isomorphic arena
involving polarized light is entirely practical. Thus, nowadays, the EPR issue
is discussed, analyzed and explored in terms of polarization phenomena.
However, note that because of Heisenberg Uncertainty, \( (x,\, p) \) do
\emph{not} commute, i.e.,  \( [\hat{x},\, \hat{p}]=i\,\hbar  \), whereas the
basis operators of polarization space \( \hat{E_{h}},\, \hat{E_{p}} \), where
\( \hat{E_{x}} \) represents an electric field in the \( x \) direction,
\emph{do} commute. That is, there is no Heisenberg Uncertainty among
polarization modes. This is a fact substantiated by QM itself;  creation and
annihilation operators for different modes of polarization, commute. Bohm's
transfer of venue moved the issue from one in which there is Heisenberg
Uncertainty into one in which there is not! Experiments in the polarization
arena can not, therefore, address the issue introduced by EPR; they, by logical
necessity, leave it unexamined. It is simply impossible to investigate
Heisenberg Uncertainty where there is none.

Because the basis operators of polarization space (a.k.a. ``qubit space'')
commute, all polarization phenomena ultimately must be describable with non
quantum principles. It is for this reason, that a classical model can explain
this experiment. Alternately, this conclusion follows forthwith, albeit with
mostly only formalistic authority, from the Optical Equivalence
Theorem.\cite{KS} 

\begin{figure*}[t]
{\par\centering\resizebox*{0.88\textwidth}{!}{\includegraphics{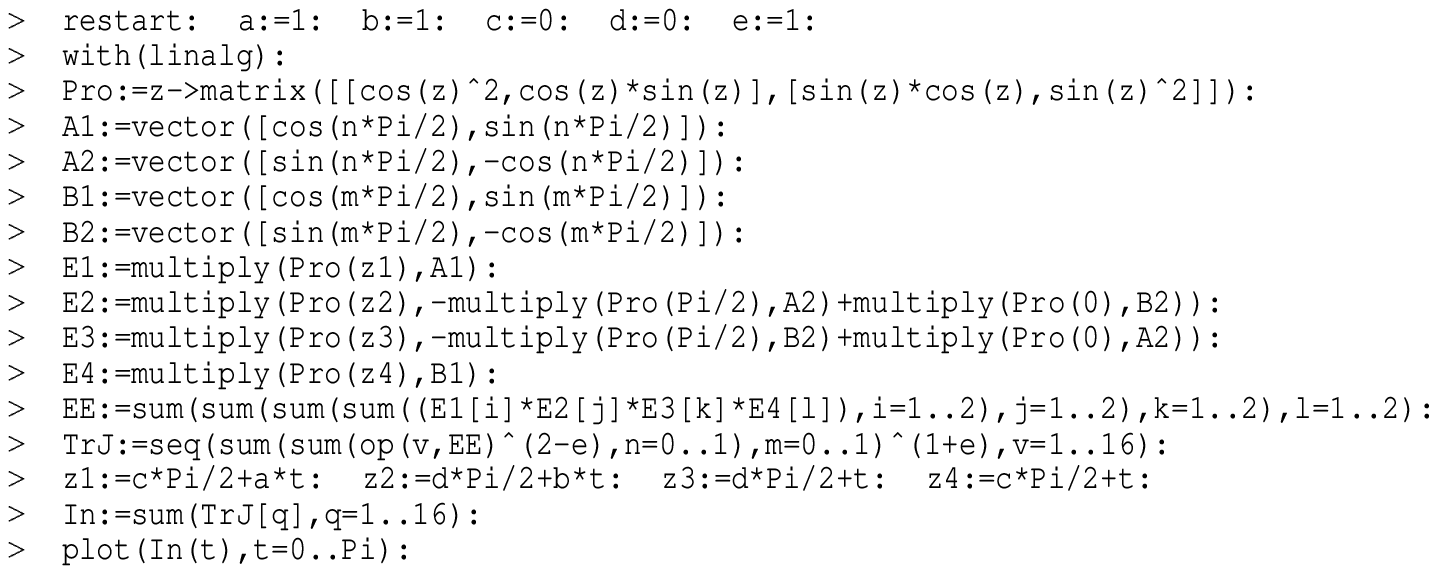}}\par}
\caption{A Maple implementation of Eq.~(\ref{e2}). The  polarizer regimes
are encoded by setting the values of \(a, b, c\)  and \(d\) to \(\pm 1,\, 0\). 
`t' is the skew angle. Setting the value of \(e\) to \(1\), adds the EM field
contribution from the various centers before `squaring;' as physics, this
accounts for ``crossover'' signals that arrive simultaneously with the
reflected signals and  interfere. Setting \(e\) to zero, on the other hand, 
prevents interference of these signals. (This works with this model because of
the peculiarities of down-conversion and a PBS;  care must be taken modeling
other setups to (in-)exclude appropriate signals before squaring.)}
\end{figure*}
\emph{}

Bohm did not justify carefully his change of venue; he simply declared spin
operators to be equivalent and sallied forth. This has been accepted,
apparently, on the grounds that, like the basis operators of phase space, \(
(x,\, p) \) , the spin operators \( (\sigma _{x},\, \sigma _{y}) \) also do not
commute. While this is indeed true, the reason is not the same. Spin operators
with discrete eigenvalues pertain to the direction in space defined by a
magnetic field. In directions transverse to the magnetic field, the expectation
values are not discrete but oscillate out of phase. In the end, the reason spin
operators do not commute would be that it is impossible to have more than one
direction for a magnetic field or precession axis at a time. Non commutation,
then,  is not a manifestation of Heisenberg Uncertainty in this case, but of
geometry, i.e., of the nature of an axis of rotation. Indeed, angular momentum
operators do not commute in classical mechanics. Similar remarks pertain
to the privileged direction defined by the ``\(\bf{k}\)'' vector of an EM wave
with respect to the transverse, i.e., polarization directions.

Entanglement is often cited as the core of QM, an idea for which perhaps
Schr\"odinger was the originator.\cite{ES35} However, if entanglement is
defined in terms of the non factorability of a wave function, as it most often
is, then it must be attributable to a correlation between subsystems.Obviously,
if a wave function factors into terms each pertaining to a separate subsystem,
\( \psi _{1}(r_1)\psi _{2}(r_2) \), then the probability density computed from
this product, \( \psi ^{*}_{1}\psi^{*}_{2}(r_1,\, r_2)\psi _{2}\psi_{1}(r_1,\,
r_2)=(\psi^{*}_{1}(r_1)\psi_{1}(r_1))(\psi^{*}_2(r_2)\psi_{2}(r_2))\), also
factors, in which case it pertains to \emph{statistically independent,} i.e.,
uncorrelated subsystems.  If the wave function does not factor, then the
probability density also will not factor and, clearly, the subsystems are
simply not uncorrelated. Correlation need not imply nonlocality; hereditary
correlation satisfies all requirements posed by the physical situation. In
experiments involving EPR style setups, including that described herein,
correlation can be vested in the `photons,' or in classical terms: the signals,
at their origin and simply carried along thereafter.

The actual problem with entanglement arises elsewhere --- with \emph{particle}
beams. A wave function describing a particle beam can not be considered simply 
as pertaining to a physical ensemble, because particles, one by one, are diffracted at
slits, for example; they suffer, seemingly, `entanglement' between Gibbsian
ensemble members. Rationalizing this phenomenon requires other
arguments\cite{Steerage}; but, the desideratum of uniformity of
principles, has lead to the mandate that radiation, too, be considered
ontologically ambiguous until the moment of measurement, even though there is
no need to do so. Its apparent digitization can be seen simply as a
manifestation of the nature of photoelectron detectors.

Of course, the tactic employed herein works for all phenomena for which there
is no Heisenberg Uncertainty, e.g., simple EPR correlations.\cite{Mepris} It
offers a decidedly less mystical interpretation of many phenomena typically
ascribed to QM. Teleportation, for example, admits a passive interpretation
involving no ``portation'' of any nature. In the above model, so-called
teleported states (1 and 4) are those which, although from separate random
sources, eventually match up, and this is signaled by an appropriate
coincidence between each's partner (2 and 3). Such an explanation is decidedly
less enchanting than that conveyed by the term ``teleportation,'' but hugely
more respectful of principles desirable for a rational explanation of the
natural world.


\begin{thebibliography}{10}
\bibitem{JWPan}J.-W. Pan, M. Daniell, S. Gasparoni, G. Weihs and A. Zeilinger,
\emph{Phys. Rev. Lett.} \textbf{86}, 4435, (2001).
\bibitem{M&W}L. Mandel and E. Wolf, \emph{Optical Coherence and Quantum Optics,} (Cambridge
University, Cambridge, 1995).
\bibitem{Klyshko}N. V. Evdokimov, D. N. Klyshko, V. P. Komolov and V. A. Yarochkin, \emph{Physics
- Uspekhi} \textbf{39,} 83 (1996).
\bibitem{DB'meester}D. Bouwmeester, et al. \emph{The Physics of Quantum Information,} (Springer,
Berlin, 2000).
\bibitem{DBohm}D. Bohm, \emph{Quantum Theory,} (Prentice-Hall, New York, 1951).
\bibitem{EPR}A. Einstein, B. Podolsky and N. Rosen, \emph{Phys. Rev.} \textbf{47}, 777 (1935).
\bibitem{KS}J. R. Klauder and E. C. G. Sudarshan, \emph{Fundamentals of Quantum
Optics}, (Benjamin, New York, 1968), p. 192.
\bibitem{ES35}E. Schr\"odinger, \emph{Naturwissenschaften,} \textbf{23,} 807, 823, 844 (1935).
\bibitem{Steerage}A. F. Kracklauer, \emph{Found. Phys. Lett.} \textbf{12} (5)
441 (1999).
\bibitem{Mepris}A. F. Kracklauer, \emph{Ann. Fond. L. deBroglie} \textbf{20} (2)
 193, (2000); quant-ph/0108057.



\end{thebibliography}
\end{document}